\documentclass[a4paper,11pt]{article}
\usepackage{pos}
\usepackage{xspace}

\newcommand{\pb}{\rm pb}

\newcommand{\GeVsq}{\ensuremath{\mathrm{GeV}^2} }
\newcommand{\GeV}{\ensuremath{\mathrm{GeV}} }

\newcommand{\Qsq}{\ensuremath{Q^{2}}\xspace}

\newcommand{\as}{\ensuremath{\alpha_{\rm s}}\xspace}

\newcommand{\tQ}{\ensuremath{\tau_Q}\xspace}
\newcommand{\tb}{\ensuremath{\tau_1^b}\xspace}

\newcommand{\pzbi}{\ensuremath{P_{z,i}^\text{Breit}}\xspace}

\title{Measurement of the 1-jettiness in deep-inelastic scattering at HERA}

\author*[\dagger, a]{J. Hessler}
\affiliation[\dagger]{On behalf of the H1 collaboration}
\affiliation[a]{Max-Planck Institute for Physics,\\
  Föhringer Ring 6, 80805 München, Deutschland}

\abstract{A first measurement of the 1-jettiness event shape observable \tb in neutral-current deep inelastic scattering is presented. 
For the measurement, the equivalence of \tb to the DIS thrust observable defined in the Breit frame is utilised.
The data were taken by the H1 experiment at HERA from 2003 to 2007 at a centre of mass energy of $\sqrt{s}=319~\GeV$. The data amount to an integrated luminosity of 351.6\,pb$^{-1}$.
The triple-differential cross sections are
presented as a function of the 1-jettiness $\tau^1_b$, the
virtuality of the exchanged boson $Q^2$ and the inelasticity of the event $y$. 
The data exhibit a sensitivity to the strong coupling constant and to resummation and hadronisation effects, as well as to the parton distribution functions of the proton. The data are compared to selected predictions. 
}

\FullConference{%
  *** The European Physical Society Conference on High Energy Physics (EPS-HEP2021), ***\\
  *** 26-30 July 2021 ***\\
  *** Online conference, jointly organized by Universität Hamburg and the research center DESY ***
}

\begin{document}
\maketitle

\section{Introduction}

\begin{figure}[thb!]
\centering
\includegraphics[height=0.38\textwidth,trim={0 0 0 0 },clip]{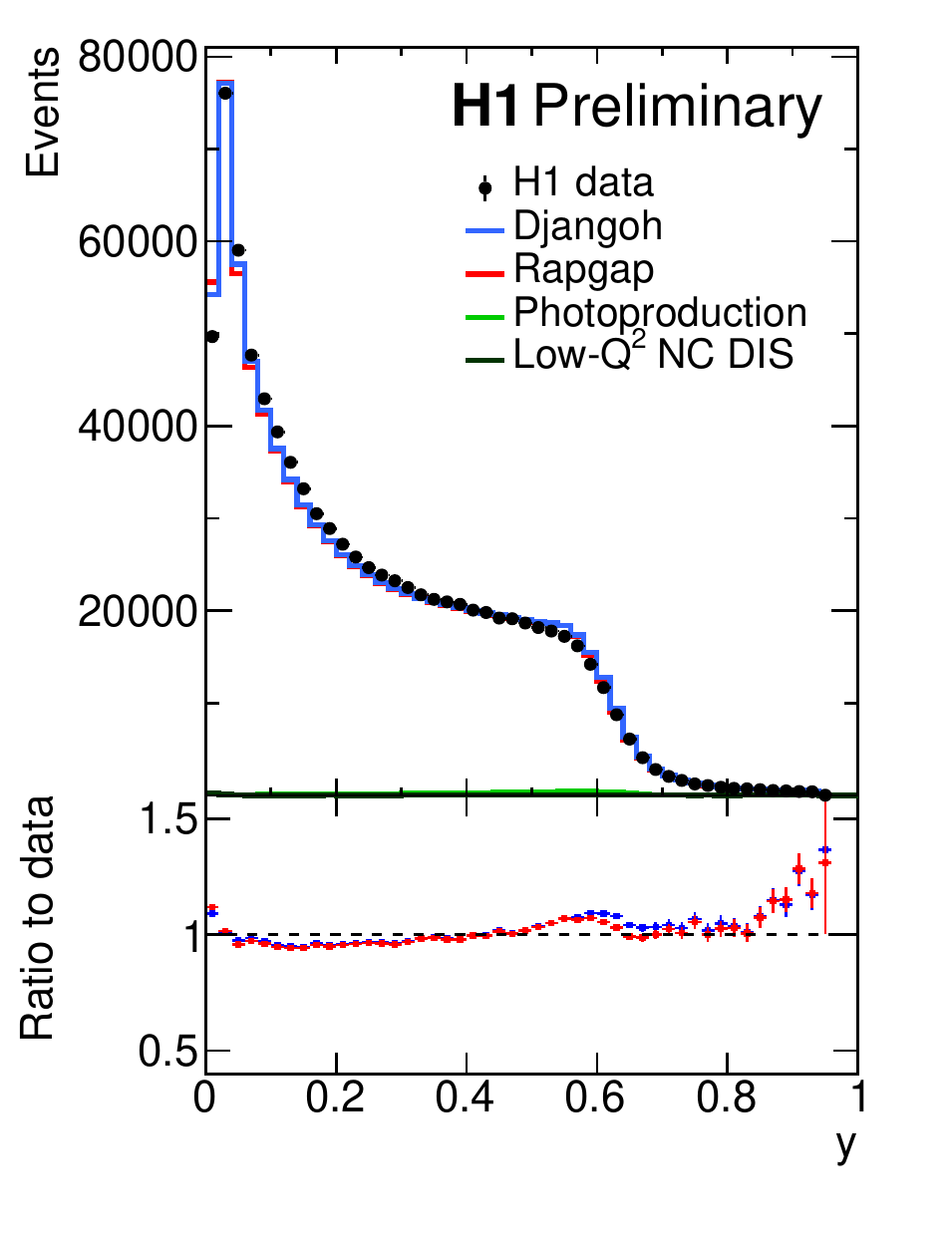}
\includegraphics[height=0.38\textwidth,trim={0 0 0 0 },clip]{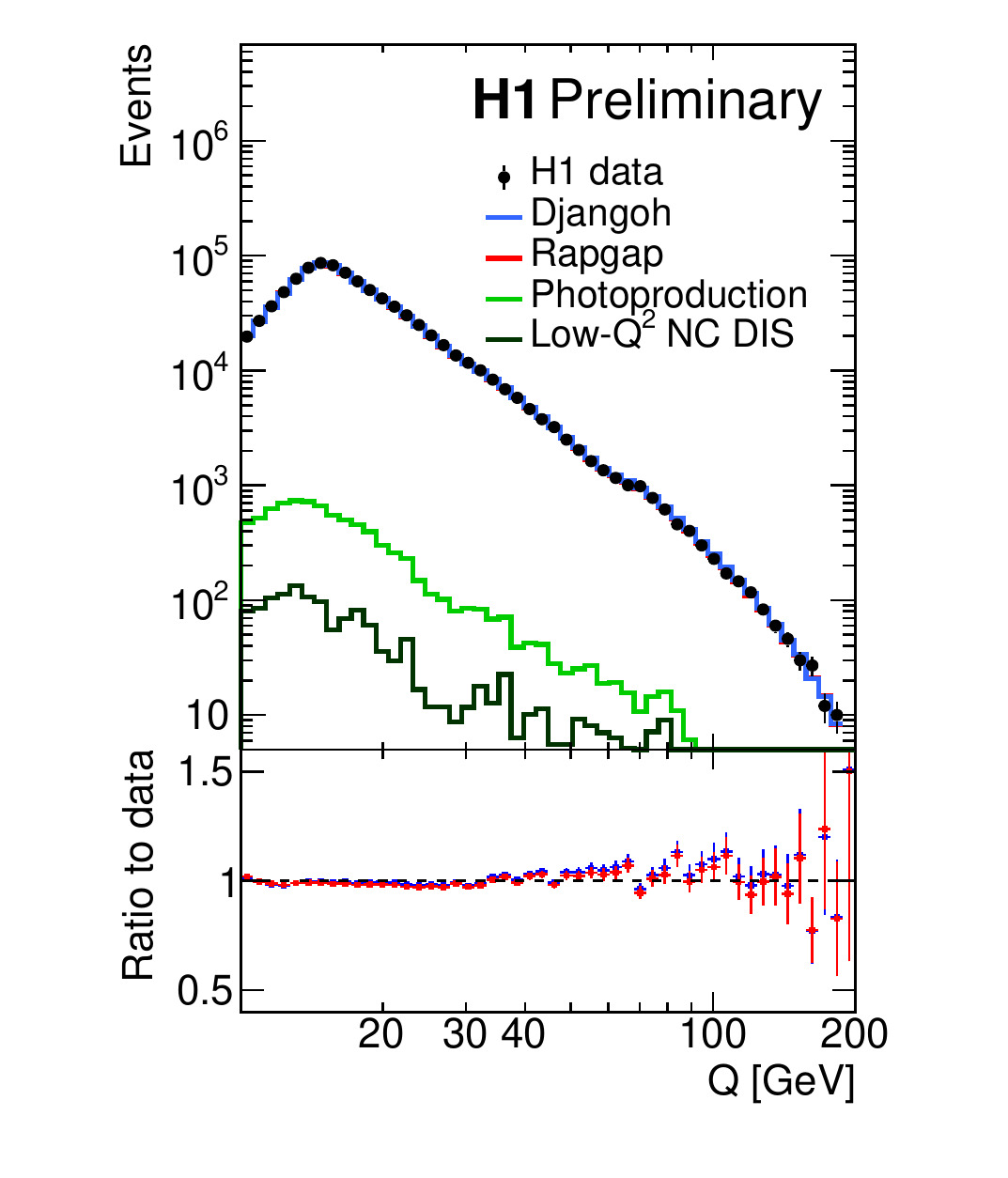}
\includegraphics[height=0.38\textwidth,trim={0 .8cm 0 0 },clip]{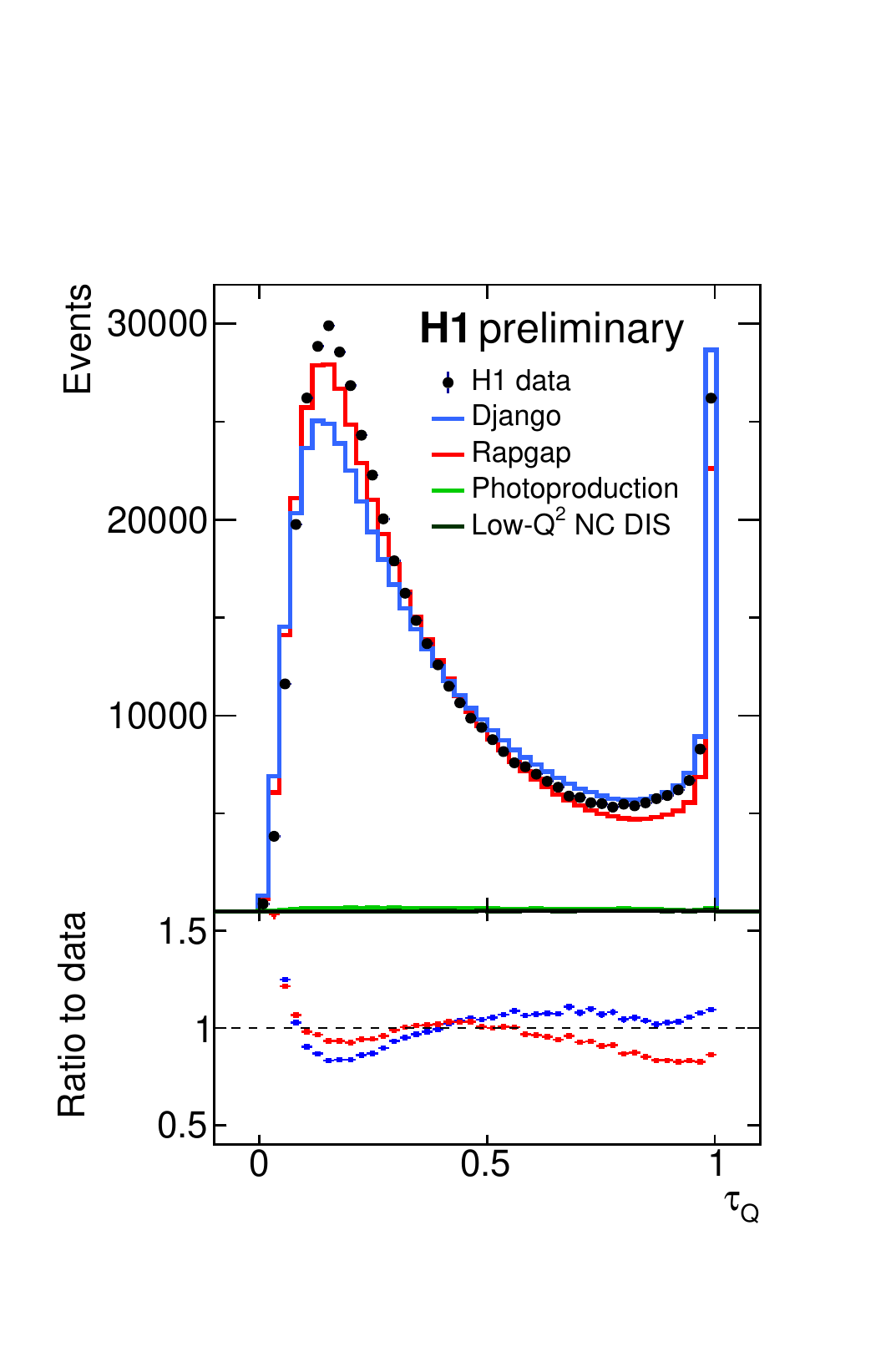}
\caption{
  Detector level distributions of $y$ (left), $Q$ (middle) and \tQ (right) of all pre-selected data. The definition of these observables is given in the equations~\eqref{eq:tauQ} and \eqref{eq:DIS}. The light green line, denote as photoproduction, represents the sum of all background processes. The lower panel shows the ratio of the event generators to the data. 
}
\label{fig:ControlPlots}
\end{figure}

With the development of future electron ion colliders, such as the EIC in Brookhaven, event shape observables in electron proton deep-inelastic scattering (DIS) experience increasing attention. 
Event shape observables have proven to have an interesting sensitivity to the strong coupling constant $\alpha_s$, to the parton distribution functions (PDFs) of the proton and to hadronisation and resummation effects.
In the past, a variety of event shape observables have been measured in neutral-current DIS at the H1 experiment ~\cite{Adloff:1997gq,Adloff:1999gn,Aktas:2005tz}
and ZEUS~\cite{ZEUS:2002tyf,ZEUS:2006vwm}.
For this analysis the 1-jettiness event shape observable \tb ~\cite{Kang:2013nha,Kang:2014qba} is of particular interest 
\begin{equation}
  \tb =  \frac{2}{\Qsq}\sum_{i\in X}\min\{xP\cdot p_i , (q+xP)\cdot p_i\}\,.
\end{equation}
The denominator $Q^2$ denotes the virtuality of the exchanged boson. It is related to the 4-momentum of the boson $q$ via $Q^2 = - q^2$.
All particles in the final state apart from the scattered lepton contribute. Those particles constitute the hadronic final state (HFS) $X$. The quantity $x$ denotes the Bjorken scaling variable, $P$ is the incoming proton and the $p_i$ are the 4-momenta of the single HFS particles.
This definition of the observable is infrared safe, allows for analytical or automatised resummation and is free of non-global logarithms~\cite{Banfi:2004yd}.
It can be predicted theoretically with high precision with tools from soft collinear effective theory (SCET).
When applying energy-momentum conservation and changing the reference frame, a different expression for \tb can be derived~\cite{Kang:2014qba}
\begin{equation}
  \tQ =  1 - \frac{2}{Q} \sum_{i\in\mathcal{H_C}}\pzbi\,.
  \label{eq:tauQ}
\end{equation}
\tQ measures the sum of longitudinal momenta of the HFS particles in the Breit frame \pzbi.
Only particles in the current hemisphere $\mathcal{H_C}$ defined by $\eta < 0$ contribute to the sum, where $\eta$ is the pseudorapidity in the Breit frame. This expression corresponds to the DIS thrust normalised to $Q/2$. Since both definitions are equivalent, the 1-jettiness \tb can be measured as \tQ.\\
The data were taken in the years 2003 to 2007 by the H1 experiment at HERA. 
Electron or positron beams were utilised at a centre of mass energy of $\sqrt{s} = 319~\GeV$. The integrated luminosity amounts to $\mathcal{L} =361~\pb^{-1}$~\cite{H1:2012wor}. \\
The events are triggered by a high-energetic cluster in the liquid argon calorimeter (LAr). The energy of the scattered lepton has to exceed $E_{e^\prime} > 11~\GeV$. This requirement ensures an efficiency above $99~\%$ for an inclusive DIS sample in the given phase space~\cite{Andreev:2014wwa}. Additional cuts are applied to suppress QED Compton events~\cite{Andreev:2014wwa}, as well as non-collision background from beam-gas interaction and beam halo~\cite{Aaron:2012qi,Andreev:2014wwa}. 
The longitudinal energy-momentum balance of the HFS and the scattered lepton is required to be  in the interval $45 < \sum_{e^\prime,i\in X} E_i-P_{z,i} < 65~\GeV$. This requirement reduces the effect of QED initial state radiation (ISR) of the beam electron.\\
Tracks and cluster hits are combined in a particle flow algorithm to reconstruct the particle candidates. 
The  energy of the HFS objects is  calibrated
with a dedicated jet-calibration sample~\cite{Kogler:2011zz}
using a neural-network based shower-classification algorithm.
The DIS kinematic variables virtuality $Q^2$, inelasticity $y$ and Bjorken-$x$ (which is required for the boost to the Breit frame)
are defined with the I$\Sigma$ reconstruction method~\cite{Bassler:1994uq,Bassler:1997tv}:
\begin{equation}
  \Qsq = \frac{E^2_{e^\prime}\sin^2\theta_{e^\prime}}{1-y}\,,~~~
  y = \frac{\Sigma}{\Sigma + E_{e^\prime} ( 1-\cos \theta_{e^\prime})}~~~\text{and}~~
 x = \frac{E_{e^\prime}}{E_p}\frac{\cos^2(\theta_{e^\prime} / 2)}{y}\,.
 \label{eq:DIS}
\end{equation}
The quantity $\Sigma$ is defined as $\Sigma = \sum_{i\in X} (E_i - P_{z,i})$, 
the polar angle of the scattered electron is denoted $\theta_{e^\prime}$, and $E_p=920\,\GeV$ is the proton beam energy. It has to be noted, that the electron beam energy does not enter the equations and hence the effect of QED ISR is small. Additionally, it was found that the I$\Sigma$ method outperforms other reconstruction methods with regard to the obtained purities.\\
In the upper panel of figure \ref{fig:ControlPlots} the detector level distributions of $y$ (left), \Qsq (middle) and \tQ (right) are shown for the pre-selected data. The data are compared to the two independent signal Monte Carlo (MC) event generators Djangoh~\cite{Charchula:1994kf} and Rapgap~\cite{Jung:1993gf}.
The overall contribution from background processes is small. 
The largest contributions are photoproduction and events migrating from different phase space regions ($\Qsq<60~\GeVsq$). Other processes (e.g. QED Compton and di-lepton production) are found to be negligible.
The kinematic variables $y$ and \Qsq are described with high precision. The models bracket the data in the \tQ distribution. The difference between Djangoh and Rapgap can be traced back to the different physics implementations. This is not a detector effect.

\section{Cross section results}
\begin{figure}[thb!]
\centering
\includegraphics[width=0.32\textwidth,trim={0 0 0 0 },clip]{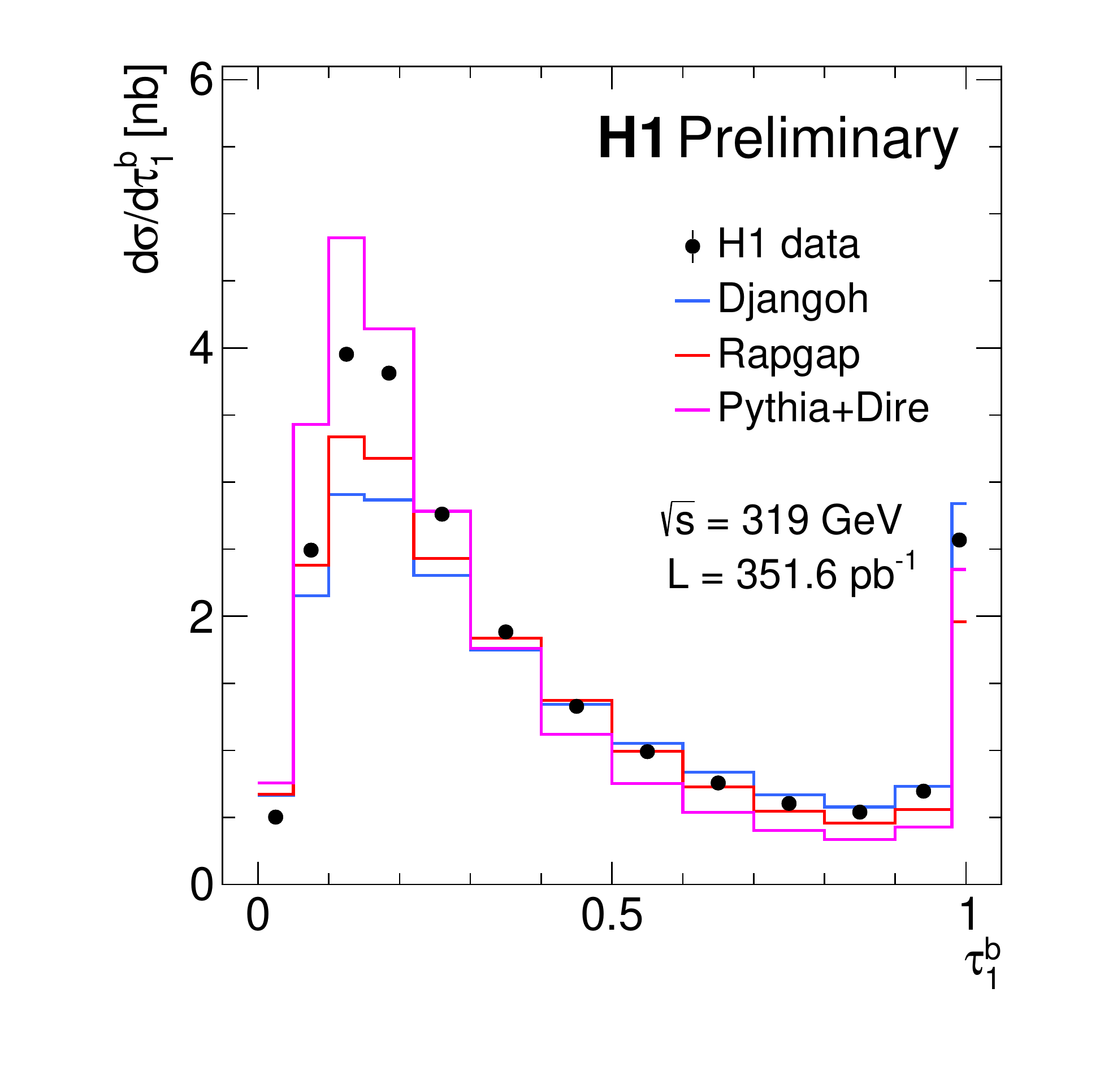}
\includegraphics[width=0.32\textwidth,trim={0 0 0 0 },clip]{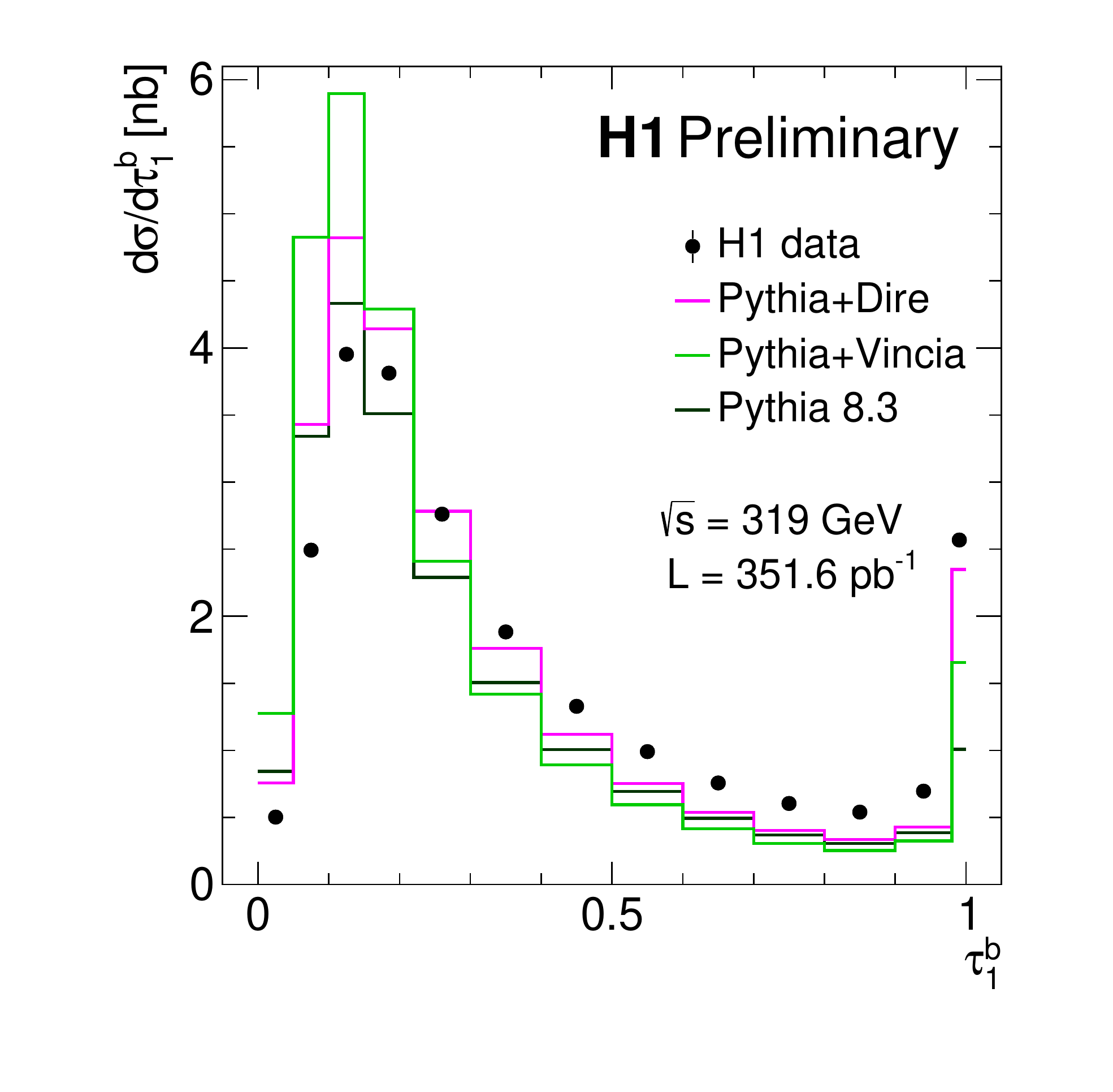}
\includegraphics[width=0.32\textwidth,trim={0 0 0 0 },clip]{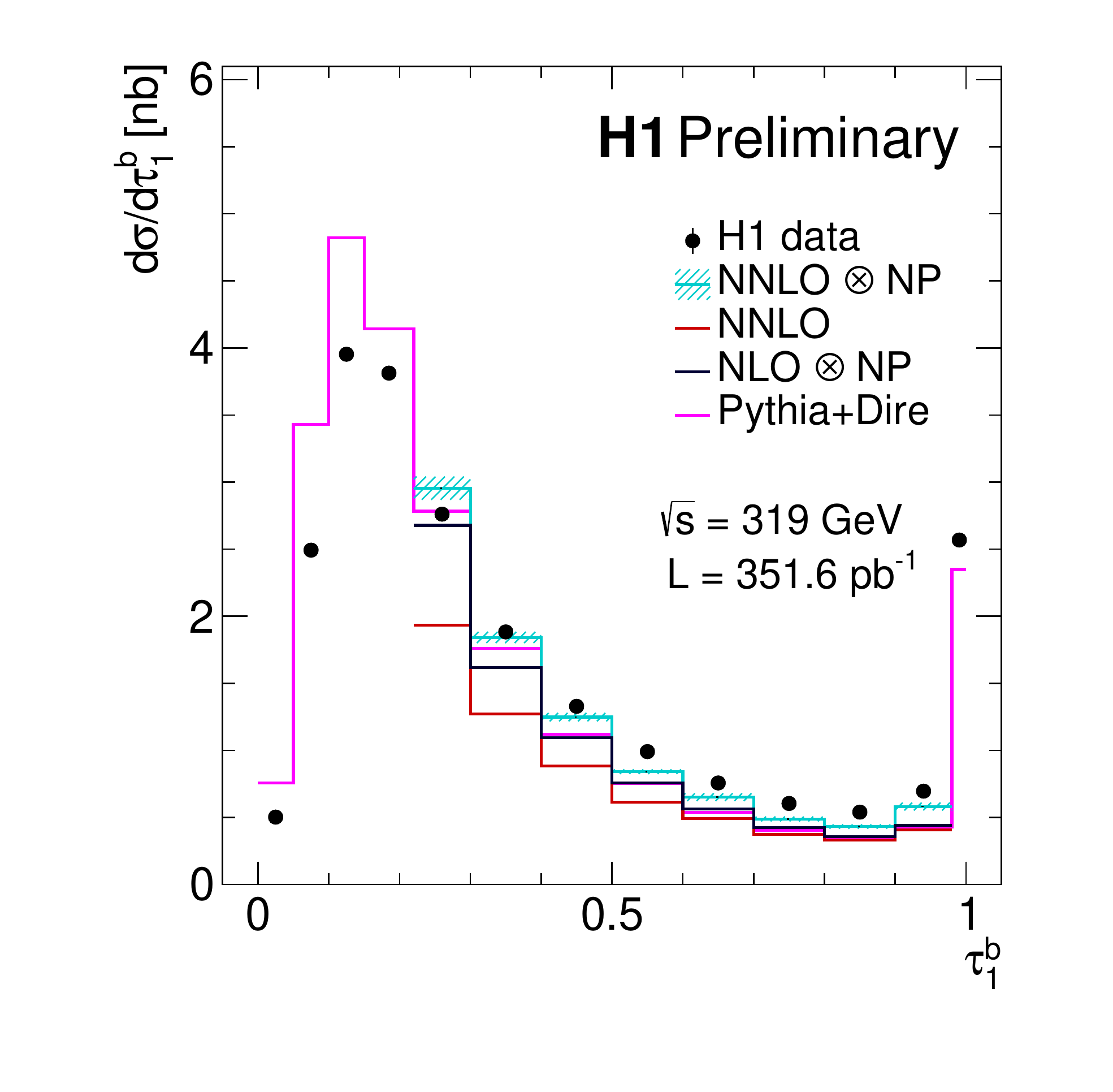}
\caption{
The differential cross section $d\sigma/d\tb$ in the kinematic region $150<\Qsq<20\,000\,\GeVsq$ and $0.2<y<0.7$.
The data are compared to the MC predictions from Djangoh and Rapgap (left), Pythia with various parton shower models (middle) and fixed order calculations (right).
}
\label{fig:1Da}
\end{figure}

The cross section $d\sigma/d\tb (\Qsq,y)$ in one measurement bin is defined as 
\begin{equation}
  \frac{d\sigma}{d\tb} = \frac{N_\text{data} - N_\text{Bkg}}{\mathcal{L} \cdot
    \Delta_\tau}\cdot c_\text{unfold}\cdot c_\text{QED}\,,
\end{equation}
where $N_\text{data}$ denotes the number of events in one bin, $N_\text{Bkg}$ denotes the estimated number of background events (processes other than high-\Qsq NC DIS), $\mathcal{L}$ is the integrated luminosity, $\Delta_\tau$ denotes the width of single \tb bins, $c_\text{unfold}$ and $c_\text{QED}$ are multiplicative factors correcting for detector and QED radiative effects, respectively.
The correction factors for the detector corrections are obtained from the two signal MC generators Djangoh~1.4~\cite{Charchula:1994kf}
and Rapgap~3.1~\cite{Jung:1993gf} in combination with a detailed detector simulation based on GEANT3~\cite{Brun:1987ma}. This correction corresponds to the bin-by-bin method, denoted as unfolding, and accounts for resolution and acceptance effects. The factors $c_\text{QED}$ are also obtained from Djangoh and Rapgap and correct for QED radiative effects of the electron. The subroutines from Heracles~\cite{Kwiatkowski:1990es} are implemented in both models for this purpose. The corrections include the emission of real photon and photonic lepton vertex corrections.\\
The results are provided as single-differential cross sections as a function of \tb in a large region in \Qsq and $y$, as well as in adjacent (\Qsq,$y$)-bins.
The latter will be denoted as triple differential cross sections. The condition on the energy of the scattered electron ($E_{e^\prime}>11~\GeV$) and the requirement that it has to be reconstructed in the LAr ($\vartheta_{e^\prime}<154^\text{o}$) define the phase space in $y$ and \Qsq. The region $y\lesssim 0.1$ (which corresponds to high $x$) is omitted from the measurement , due to limited acceptance and resolution in the very forward detector region.
For the single differential measurement this translates to the phase space $150 < \Qsq < 20~000\GeVsq$ and $0.2 < y < 0.7$ and for the triple differential measurement $150 < \Qsq < 20~000\GeVsq$ and $0.1 < y < 0.9$. \\
The integrated luminosity is associated with a systematical uncertainty of $2.7~\%$~\cite{H1:2012wor}. This is the dominant uncertainty of the measurement. The other sources taken into account are the energy uncertainty of the scattered lepton~\cite{H1:2012qti} and the polar-angle position of the LAr with respect to the Central Tracking Detector~\cite{H1:2012qti}.
The latter is studied for the lepton and HFS particles separately. 
The energy of single HFS objects is assigned after a dedicated jet energy calibration~\cite{Kogler:2011zz,Andreev:2014wwa}.
By varying the energy of the single HFS particles by $1~\%$, the uncertainties following from the calibration procedure are obtained.\\
The measured cross sections are compared to various predictions.
The DIS MC event generators Djangoh~1.4~\cite{Charchula:1994kf}
and Rapgap~3.1~\cite{Jung:1993gf} are used in combination with the Lund string fragmentation model~\cite{Andersson:1983ia,Sjostrand:1994kzr} with the ALEPH tune~\cite{ALEPH:1996oqp} and the CTEQ6L PDF~\cite{Pumplin:2002vw}.
The data are corrected for higher-order QED radiation effects. Thus these are not included in these predictions. \\
The MC event generator Pythia~8.303~\cite{Sjostrand:2014zea,Pythia83} is used together
with three different implementations for the parton shower: the
`default' 
shower, the Vincia parton shower~\cite{Giele:2007di,Giele:2011cb,Giele:2013ema,Fischer:2016vfv} and the Dire~\cite{Hoche:2015sya,Hoche:2017iem,Hoche:2017hno} parton shower. For all three models the NNPDF3.1 PDF set~\cite{NNPDF:2017mvq} is used with $\as = 0.118$.
The Pythia~8.3 default for hadronisation is used~\cite{Pythia83}.\\
Next-to-next-to-leading order (NNLO) predictions in perturbative QCD for the process $ep\to e+2\text{jets}+X$ are obtained with the program NNLOJET~\cite{Ridder:2016nkl,Currie:2017tpe,Currie:2016ytq,Gehrmann:2019hwf} with $\mu_R = \mu_F =Q$ and the NNPDF3.1~\cite{NNPDF:2017mvq} PDF set.
Multiplicative corrections for hadronisation effects are obtained from Pythia~8.3.
These NNLO predictions are valid only in the region $\tb\gtrsim0.22$ and $\tb\neq1$. NLO predictions for the same region are displayed for comparison.\\
Figure \ref{fig:1Da} shows the single differential corss section results. 
The statistical and systematical uncertainties are commonly smaller than the marker size and are not shown here.
The data show a distinct peak in the region $\tb \lesssim 0.3$ (denoted as peak region) and a decreasing cross section towards higher values of \tb (denoted as tail region). The peak region consists of DIS one jet events and is sensitive to resummation and hadronisation effects. The tail region is populated by events with additional hard radiation. In DIS an event configuration with empty current hemisphere is possible. If the struck parton emits hard QCD radiation, both partons can be kicked into the beam hemisphere. This type of event topologies populates the last bin $0.98 \leq \tb \leq 1$.\\
The data are compared to the classical models Djangoh and Rapgap (left), to Pythia predictions (middle) and to fixed order calculations (right). The Pythia+Dire prediction is shown in all three plots to enable a better comparison.
Rapgap and Djangoh underestimate the $\tb \lesssim 0.3$ region (denoted as peak region) but provide a satisfactory description of the $\tb > 0.3$ region (denoted as tail region). The peak region is very sensitive to resummation and hadronisation effects. None of the Pythia parton shower implementations succeed in describing the data.
The Pythia predictions behave similarly in the tail region and underestimate the data. The NNLO calculations provide a reasonable description of the data in the region of validity. However, the hadronisation corrections can get sizeable. \\

\begin{figure}[thb!]
\centering
\includegraphics[width=0.82\textwidth,trim={0 0 0 0 },clip]{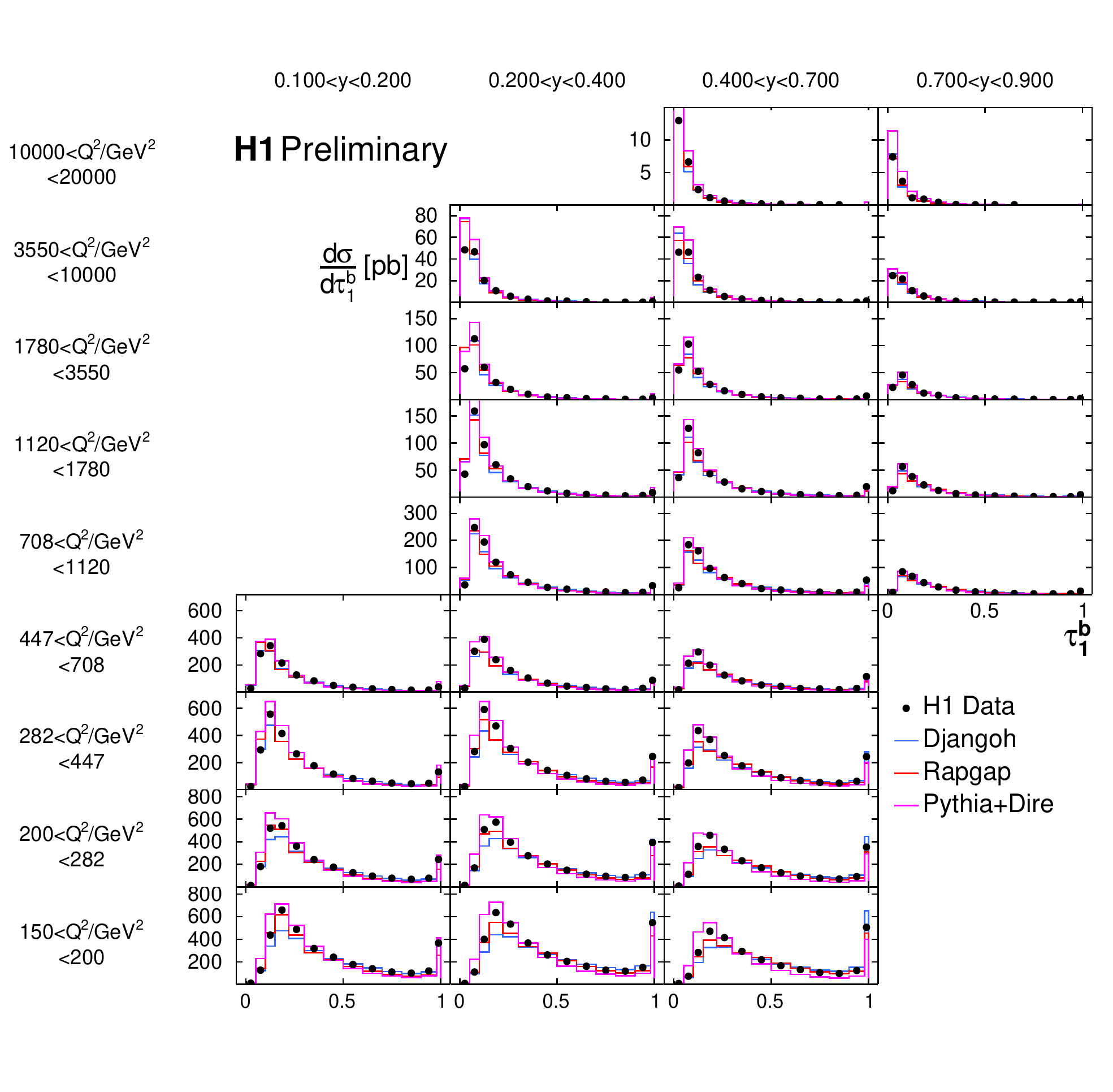}
\caption{
  The differential cross section $d\sigma/ d\tb$ for adjacent regions in \Qsq\ and $y$.
  Every panel displays the differential cross
  section $d\sigma/ d\tb$  in a given phase space which is indicated on the left and top for \Qsq and $y$, respectively. 
  The data are compared to predictions from the Djangoh and Rapgap, where QED radiative effects were switched off.
  Predictions from Pythia8.3 using the Dire parton shower model are further displayed.
}
\label{fig:xs3Dold}
\end{figure}

The triple differential cross sections in adjacent bins of \Qsq and $y$ are presented in figure \ref{fig:xs3Dold}. The \Qsq and $y$ bins are indicated on the left and at the top, respectively. Increasing \Qsq shifts the peak region towards lower \tb values and lowers the tail region. At high \Qsq the momentum of the Born level DIS-jet increases and the probability of hard QCD radiation is reduced. Increasing $y$ (which corresponds to lowering $x$) enhances the $\delta$-peak in the last bin. The event topology with empty current hemisphere is only present in events with $x$ at least smaller than 0.5~\cite{Kang:2014qba}.\\
Djangoh performs best in describing the data, while Rapgap underestimates the high \tb region at low $y$.
The Pythia+Dire prediction is similar to Rapgap at low-$y$
but overestimates the data at low \tb, while underestimating the tail region. The comparison to the other predictions can not be shown here, due to a lack of space. They are displayed elsewhere \cite{TauQ:2021}.

\section{Summary and conclusion}
A first measurement of the 1-jettiness event shape observable \tb in deep-inelastic electron proton scattering at HERA is presented. The data were taken with the H1 experiment. The equivalence of \tb to the classical thrust observable \tQ was employed. The cross sections were presented in a single and triple differential manner in the phase space $150 < \Qsq < 20~000\GeVsq$ and $0.2 < y < 0.7$ and $0.1 < y < 0.9$, respectively. The data are compared to modern Pyhtia 8.3 predictions with various parton shower models, to fixed order NNLO calculations and to the MC models Rapgap and Djangoh. Only the latter provide a satisfactory description of the data. The measurement is sensitive to the strong coupling constant, to the proton PDF and to resummation and hadronisation effects. It will become valuable for improving multi-purpose MC event generators and to test QCD at highest precision.

\bibliography{tau1b_prelim}

\end{document}